# Physics-informed Neural Networks Enable High Fidelity Shear Wave Viscoelastography across Multiple organs

Ziying Yin[1], Yuxuan Jiang[1], Yuxi Guo[1], Jiayi Pu[1], Shiyu Ma[1], Guo-Yang Li[2], Yanping Cao[1,*]

[1]Institute of Biomechanics and Medical Engineering, AML, Department of Engineering Mechanics, Tsinghua University, Beijing 100084, China.

[2]Department of Mechanics and Engineering Science, College of Engineering, Peking University, Beijing 100871, China.

*Corresponding author: caoyanping@tsinghua.edu.cn (Y. P C.);

# Abstract

Tissue viscoelasticity has been recognized as a crucial biomechanical indicator for disease diagnosis and therapeutic monitoring. Conventional shear wave elastography techniques depend on dispersion analysis and face fundamental limitations in clinical scenarios. Particularly, limited wave propagation data with low signal-to-noise ratios, along with challenges in discriminating between dual dispersion sources stemming from viscoelasticity and finite tissue dimensions, pose great difficulties for extracting dispersion relation. In this study, we introduce SWVE-Net, a framework for shear wave viscoelasticity imaging based on a physics-informed neural network (PINN). SWVE-Net circumvents dispersion analysis by directly incorporating the viscoelasticity wave motion equation into the loss functions of the PINN. Finite element simulations have revealed that SWVE-Net allows for the quantification of viscosity parameters within a wide range (e.g., 0.15 – 1.5 Pa·s). Remarkably, it can achieve this even for samples as small as a few millimeters, where substantial wave reflections and dispersion take place. Ex vivo experiments have demonstrated the broad applicability of SWVE-Net across various organ types, including the brain, liver, kidney, and spleen, each with distinct viscoelastic characteristics. In in vivo human trials on breast and skeletal muscle tissues, SWVE-Net provides a reliable assessment of viscoelastic properties of these living tissues with the ratios of standard deviations to the mean values being less than 15%. This result highlights the method's robustness under real-world imaging constraints. SWVE-Net overcomes the fundamental limitations of conventional elastography and enables reliable viscoelastic characterization in situations where traditional methods fall short. Therefore, it holds promising applications such as in grading the severity of hepatic lipid accumulation, detecting myocardial infarction boundaries, and distinguishing between malignant and benign tumors.

# Introduction

Biological soft tissues ubiquitously demonstrate time-dependent viscoelastic responses under various mechanical stimuli [1-4]. Accumulating evidence indicates that tissue viscoelasticity is not only influenced by physiological states, but also undergoes dynamic alterations during disease pathogenesis [5, 6]. In previous studies, significant variations in viscoelastic parameters have been reported in multiple pathological conditions, including hepatic steatosis (fatty liver disease) [7], ischemic myocardium following acute myocardial infarction [8], and solid tumor development [6]. Specifically, diseased tissues exhibit measurable alterations in their viscoelastic properties when compared to healthy counterparts, such as changes in characteristic stress relaxation time and frequency-dependent storage/loss moduli [9]. These findings collectively demonstrate that quantitatively probing tissue viscoelastic metrics in vivo may provide diagnostic biomarkers in clinics, for instance, for grading hepatic lipid accumulation severity, detecting myocardial infarction boundaries, discriminating malignant from benign tumors, and monitoring treatment response in hepatitis disorders.

In clinical settings, shear wave elastography (SWE) seems to stand as the sole approach for quantifying the viscoelastic parameters of both superficial and deep tissues in vivo. This is achieved by utilizing either ultrasound or magnetic resonance imaging to monitor the propagation of shear waves [10-16]. Tissue viscosity gives rise to wave dispersion, meaning that the shear wave velocity (SWV) varies in accordance with the wave frequency. Consequently, the majority of methods reported in the literature and essentially all the methods currently used in clinics center on the determination of the dispersion relation, namely the relationship between SWV and frequency [17]. This relation is then employed to deduce the viscoelastic parameters of tissues. To extract the dispersion curve from wave motion signals, the two-dimensional fast Fourier transform (2D FFT) method has been extensively applied [18-20]. Nonetheless, the limited amount of data accessible from in vivo tests presents difficulties for this classical method. In such circumstances, the two-point method demonstrates its utility. It capitalizes on the wave motion information measured at two lateral locations and uses wavelet transformation analysis to derive the dispersion curves [21]. It is important to note that both viscoelasticity and finite dimensions can lead to wave dispersion. As a result, these SWE methods based on the wave dispersion information are ineffective for inferring the viscoelastic properties of small-sized soft tissues (for example, when tissue typical size is comparable to or even smaller than the wavelength). This is because they are incapable of separating the viscosity-induced dispersion from the structure-induced dispersion within the dispersion curve.

To tackle the aforementioned challenge in the quantitative assessment of tissue viscoelastic parameters, this study reports a physics-informed neural network [22] (PINN) enhanced shear-wave visco-elastography method (SWVE-Net). This method constructs an end-to-end mapping between wave motion information and tissue viscoelastic properties. The architecture of SWVE-Net is composed of three core components. Firstly, the current method is free from dispersion curves. It employs the particle velocities that are measurable in ultrasound imaging as inputs, rather than shear wave velocities. Secondly, SWVE-Net directly incorporates viscoelastic wave equations into the loss function. This allows for the simultaneous solution of forward modeling and the inference of model parameters, without the need for extensive training datasets. Thirdly, SWVE-Net facilitates the utilization of multi-source data in the inversion of viscoelastic parameters, effectively handling the ill-posed nature of the inverse problem. This is especially important for exploring the viscoelastic parameters of small-sized tissues. In this case, reflection wave signals in the tested subject from different directions can be utilized as multi-source data. In contrast, in dispersion-curve-based methods, reflection waves pose great difficulties for the quantification of the dispersion curve.

To validate the performance of SWVE-Net, numerical experiments were first carried out. A finite element (FE) model was developed to simulate viscoelastic material behavior under propagating shear waves. For computational efficiency and illustrating the use of SWVE-Net, the Kelvin-Voigt constitutive model was implemented through a user material subroutine (UMAT) in ABAQUS [23], with two distinct viscosity coefficients considered to demonstrate the method's effectiveness across different viscoelastic regimes. The vertical particle velocity fields extracted from simulations were used to train SWVE-Net, enabling direct viscoelastic property inference without requiring dispersion analysis. Additionally, a separate FE model replicating the typical dimensions of small-sized biological tissues was constructed to evaluate SWVE-Net's performance in complex wave reflection scenarios.

To further demonstrate SWVE-Net's practical utility, we performed comprehensive *ex vivo* and *in vivo* experiments. Shear waves were generated in target tissues using focused acoustic radiation force and tracked via ultrafast ultrasound imaging at 10,000 frames per second. Utilizing the spatiotemporal particle velocity field of shear wave propagation, SWVE-Net achieved simultaneous reconstruction of both elastic modulus and viscosity parameters, demonstrating consistent stability and robust performance. Initial validation using murine liver specimens confirmed the method's capability for small-tissue characterization (e.g., the tissue dimension is as small as several millimeters). Subsequent ex vivo tests on spleen, kidney, and brain tissues yielded characteristic time constants ($10^{-4}$ s/0.1 ms) consistent with literature values. Finally, successful *in vivo* implementations on human male biceps and female breast tissues confirmed the method's clinical potential for noninvasive viscoelastic assessment.

## Results

**A dispersion-relation-free framework for shear wave viscoelasticity imaging via physics-informed neural networks (SWVE-Net)**

Here we propose SWVE-Net for viscoelastic soft material, which is an extension of our previous study, a physics-informed neural network (PINN)-based SWE method, termed SWENet [24]. The deep neural network is set up to describe the full shear wave field ($\psi$, $p$) in SWVE-Net. It takes the spatial coordinates and time as inputs, and the outputs of neural network are introduced into the residual calculations of governing equations and observational data, as shown in Fig. 1.

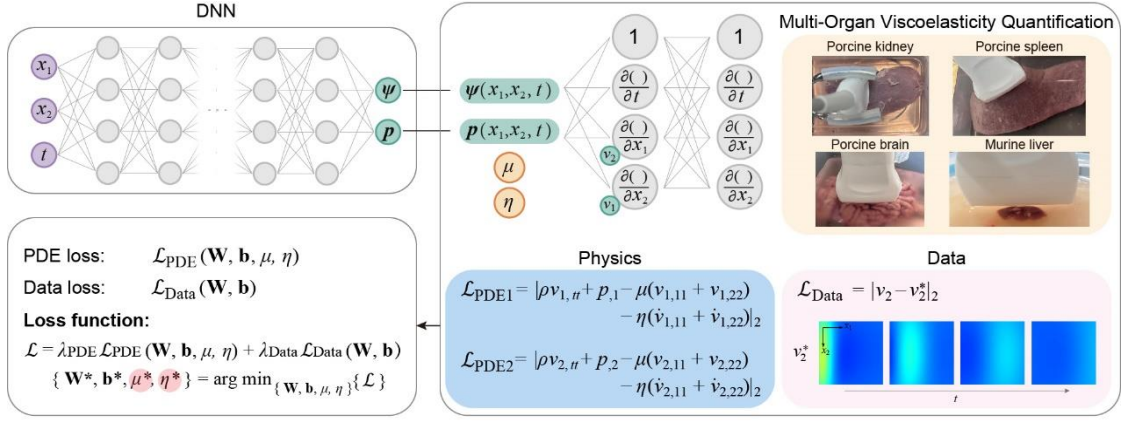

**Fig.1. Schematic of SWVE-Net for viscoelastic shear wave inversion.** SWVE-Net is a physics-informed neural network (PINN) designed to simultaneously reconstruct full-field shear wave propagation and infer viscoelastic parameters (shear modulus µ and viscosity η). The network takes spatial-temporal coordinates $(x, y, t)$ as input and outputs wavefield components $\psi$ and $p$, which are then used to compute the residuals of the governing viscoelastic wave equations. The total loss combines the physics-based residual loss and the data mismatch loss, weighted by hyperparameters. The framework enables quantitative viscoelastic characterization across multiple tissue types, including kidney, spleen, brain, liver, muscle and breast tissue.

The residuals of the governing equations (Eq. (8), see *Materials and Methods* for details) are then regularized and incorporated into the loss function as physics-informed component $\mathcal{L}_{PDE}$ (= $\mathcal{L}_{PDE1} + \mathcal{L}_{PDE2}$), where

$$\begin{cases} \mathcal{L}_{\text{PDE1}} = \left| \mu \left( \frac{\partial^2 v_1}{\partial x_1^2} + \frac{\partial^2 v_1}{\partial x_2^2} \right) + \eta \left( \frac{\partial^2 \dot{v}_1}{\partial x_1^2} + \frac{\partial^2 \dot{v}_1}{\partial x_2^2} \right) - p_{,1} - \rho \frac{\partial^2 v_1}{\partial t^2} \right|_2 \\ \mathcal{L}_{\text{PDE2}} = \left| \mu \left( \frac{\partial^2 v_2}{\partial x_1^2} + \frac{\partial^2 v_2}{\partial x_2^2} \right) + \eta \left( \frac{\partial^2 \dot{v}_2}{\partial x_1^2} + \frac{\partial^2 \dot{v}_2}{\partial x_2^2} \right) - p_{,2} - \rho \frac{\partial^2 v_2}{\partial t^2} \right|_2 \end{cases}, \quad (1)$$

and '$|\ |_2$' denotes the $L_2$ norm. The difference between the experimental data $v_2^*$ and the neural network predictions $v_2$ constitutes the data-driven component of the loss function as

$$\mathcal{L}_{\text{Data}} = |v_2 - v_2^*|_2. \quad (2)$$

The hyperparameters $\lambda_{PDE}$ and $\lambda_{Data}$ balance the contributions of the physics-informed and data-driven terms. By minimizing the total loss $\mathcal{L}$ $(=\sum_{i=1}^{M} \lambda_{PDE}\mathcal{L}_{PDE} + \sum_{i=1}^{M} \lambda_{Data}\mathcal{L}_{Data})$, SWVE-Net achieves full shear wave inversion, allowing it to infer the viscoelastic property $\mu$ and $\eta$ simultaneously.

**Finite element simulations validate SWVE-Net**

We first assessed the performance of SWVE-Net on numerical data generated from a two-dimensional finite element (FE) model, as shown in Fig. 2. Viscoelastic soft materials with the identical shear moduli but distinct viscosity parameters (i.e., viscos parameter η in Kelvin model varies from 0.15Pa·s to 1.5Pa·s) were investigated; it is noteworthy that this viscosity regime encompasses most soft tissues.

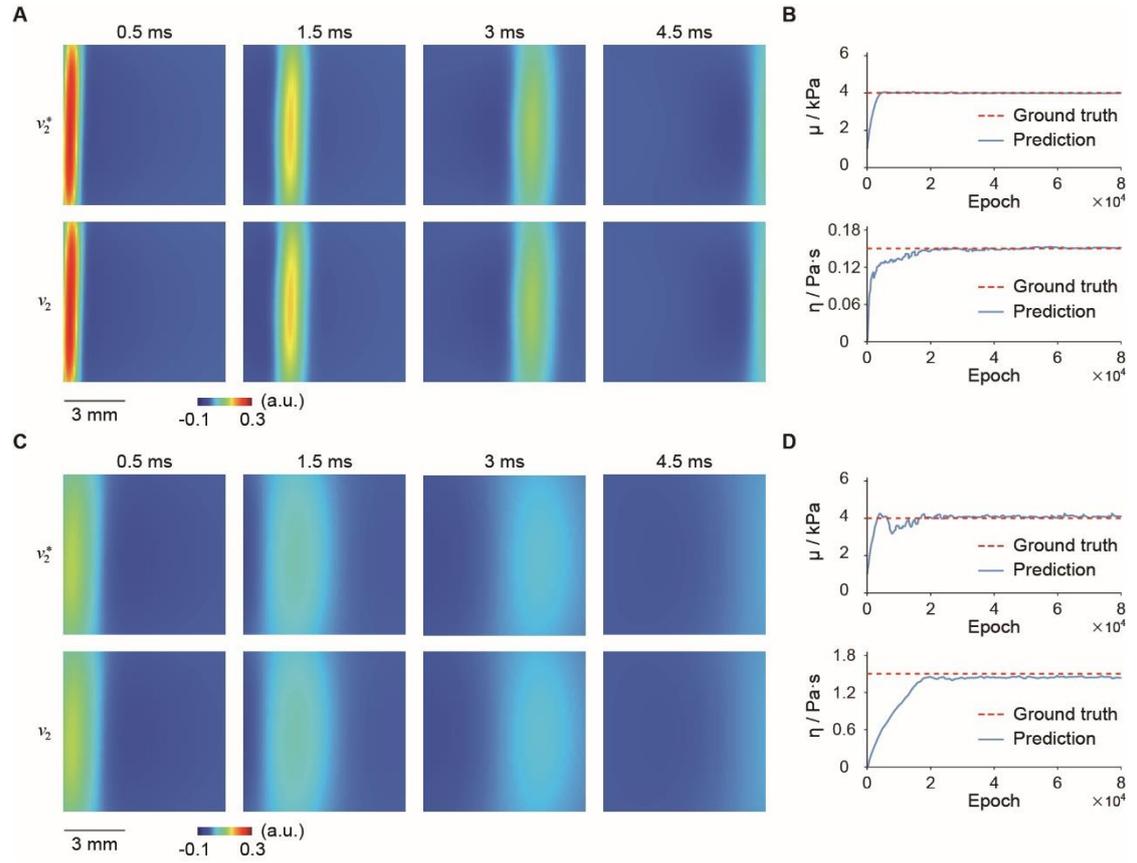

**Fig.2. Numerical validation of SWVE-Net on viscoelastic media with varying viscosity.** The outcomes of two representative examples are presented: (**A**) and (**B**) pertain to the case with low viscosity (η = 0.15 Pa·s), while (**C**) and (**D**) correspond to those with high viscosity (η = 1.5 Pa·s). In both cases, the shear modulus remains the same (μ = 4.0 kPa). For (**A**) and (**C**), snapshots of the vertical particle velocity $v_2^*$ derived from finite element (FE) simulations are shown in the first row, and the corresponding predictions of vertical particle velocity $v_2$ made by the SWVE-Net are presented in the second row. These snapshots are captured at 0.5, 1.5, 3.0, and 4.5 ms. As for (**B**) and (**D**), they depict the convergence curves of the predicted shear modulus μ and viscosity η

throughout the training epochs.

In Figs. 2A and 2C, the first row depicts the vertical component $v_2^*$ of the particle velocity field derived from finite element (FE) simulations, which served as the simulated experimental data input to SWVE-Net. The second row illustrates the network's prediction of the vertical component $v_2$ of the particle velocity after 80,000 training iterations. Notably, distinct differences in shear wave attenuation behavior emerge for the two materials with distinct viscosity parameter, with higher viscosity inducing more pronounced wave damping. Despite these disparities, Figs. 2A and 2C show that SWVE-Net successfully captured the salient features of the wave fields for both materials.

Figures 2B and 2D illustrate the convergence curves for the shear modulus and viscosity coefficient of the Kelvin-Voigt model during the inference process utilizing the SWVE-Net. As demonstrated, the SWVE-Net successfully recovered the viscoelastic parameters with remarkable accuracy, exhibiting relative errors below 5% when compared to the ground truth values assigned in the finite element simulations. Importantly, this inversion was achieved without relying on the determination of shear wave velocities and the extraction of dispersion curve. The accuracy of the inferred viscoelastic parameters demonstrates the efficiency and robustness of the proposed method.

We further assessed the performance of SWVE-Net on a small-scale model, which was designed to approximate the geometry of a cross-section of a murine liver as visualized under ultrasound imaging. In this scenario, both viscosity-induced dispersion and structure-induced dispersion are present, and decoupling them is by no means straightforward. The geometry of the model is depicted in Fig. 3A, where the locations of the acoustic radiation force (ARF) excitations and the signal acquisition region employed for mechanical property reconstruction are also denoted. The particle velocity field derived from finite element simulations is presented in Fig. 3B. Owing to the small dimension, prominent wave reflections from the boundaries were detected, leading to a complex shear wave field. Such complex wave reflections present a substantial challenge to conventional inversion methods based on dispersion analysis;

whereas the reflected waves may sever as multi-source data in SWVE-Net, facilitating the inversion of viscoelastic parameters.

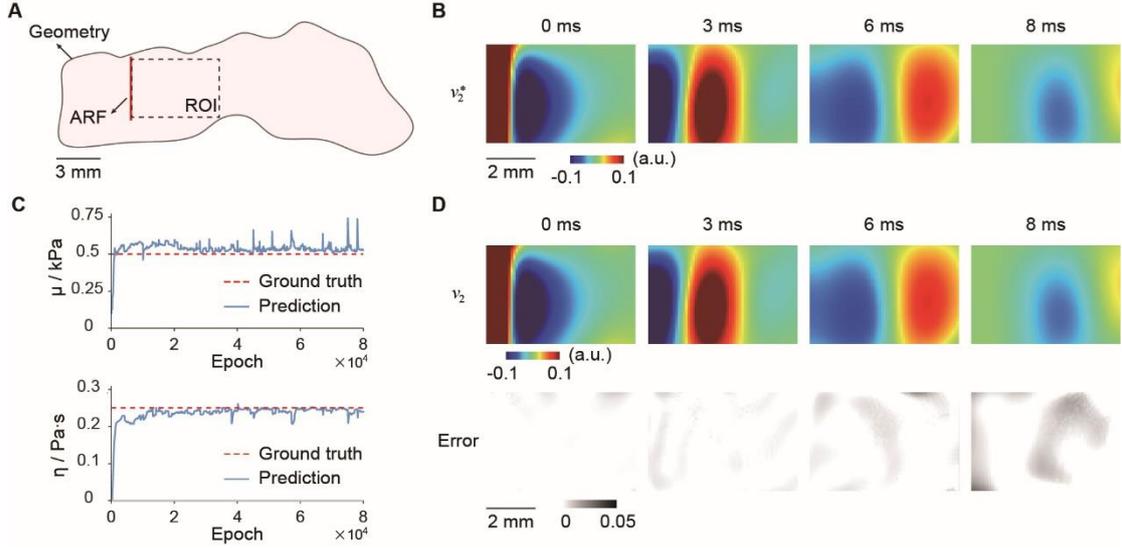

**Fig.3. Numerical validation of SWVE-Net on a small-scale murine liver model. (A)** Illustration of FE model with locations of ARF excitations (red line) and region of interest (ROI) for inversion of viscoelastic parameters (dashed black square). **(B)** Vertical particle velocity $v_2^*$ in the ROI given by FE simulations. **(C)** Convergence of viscoelastic parameters predicted by SWVE-Net (blue curve) and the ground truth values (red dashed line). **(D)** First row, SWVE-Net predictions of vertical particle velocity $v_2$; second row, relative errors between predictions and ground truth values, defined as $|(v_2 - v_2^*)/ \max(v_2^*)|$, for each snapshot.

The particle velocities observed within the region of interest were employed as simulated experimental data to train the SWVE-Net model. The wave field predicted by the network is depicted in Fig. 3D, showcasing remarkable agreement with the simulated wave patterns, including intricate reflection effects. Importantly, SWVE-Net successfully recovered the underlying viscoelastic parameters within the region of interest with relative errors not exceeding 5% compared to the ground truth values. This numerical experiment indicates that SWVE-Net is capable of robustly learning complex shear wave fields and accurately inverting the viscoelastic parameters, even in small tissue samples measuring just a few millimeters in typical dimension.

For comparison, we also performed conventional dispersion curve analysis on the same dataset using a two-dimensional spatiotemporal Fourier transform (2D-FFT) approach. Specifically, the vertical component of particle velocity along a lateral sampling line at the center of the region of interest was extracted and transformed into the frequency–wavenumber (f–k) domain. The ridge of maximum spectral amplitude in the f–k map was tracked to derive the frequency-dependent phase velocity curve in the frequency range of 0~320 Hz, which was then fitted using the theoretical solution based on the Kelvin–Voigt model to infer the viscoelastic parameters. The estimated shear modulus and viscosity were 0.44 kPa and 0.36 Pa·s, corresponding to relative errors of –12% and +44%, respectively.

While the dispersion-based method provided a reasonably accurate estimate of the shear modulus, its estimation of viscosity was considerably less reliable, due in part to sensitivity to limited dimension and structural dispersion. In contrast, SWVE-Net leverages the full spatiotemporal wavefield and incorporates the governing viscoelastic wave equations directly into a physics-informed loss function, enabling robust parameter inference even under complex wave propagation conditions. These results underscore the practical advantage of SWVE-Net over conventional dispersion-based approaches, particularly in small, reflection-prone samples such as the murine liver.

**SWVE-Net quantifies the viscoelastic properties of a murine liver with dimension as small as several millimeters**

In conjunction with the numerical validation, we carried out ex-vivo experiments to comprehensively evaluate the performance of SWVE-Net under actual measurement conditions, especially in small-organ imaging scenarios. Figure 4A presents photographs of a murine liver sample captured from multiple perspectives, indicating that the thickness of the sample was even less than 1 cm. To reduce the artifacts originating from the container, the liver samples were positioned on a gelatin block during ultrasound data acquisition, as depicted in Figure 4B. Moreover, the physical dimensions of the sample were remarkably small when compared to the L9-4 linear array probe, which demonstrates the additional challenges in both wave excitation and wave motion tracking in small sample measurements.

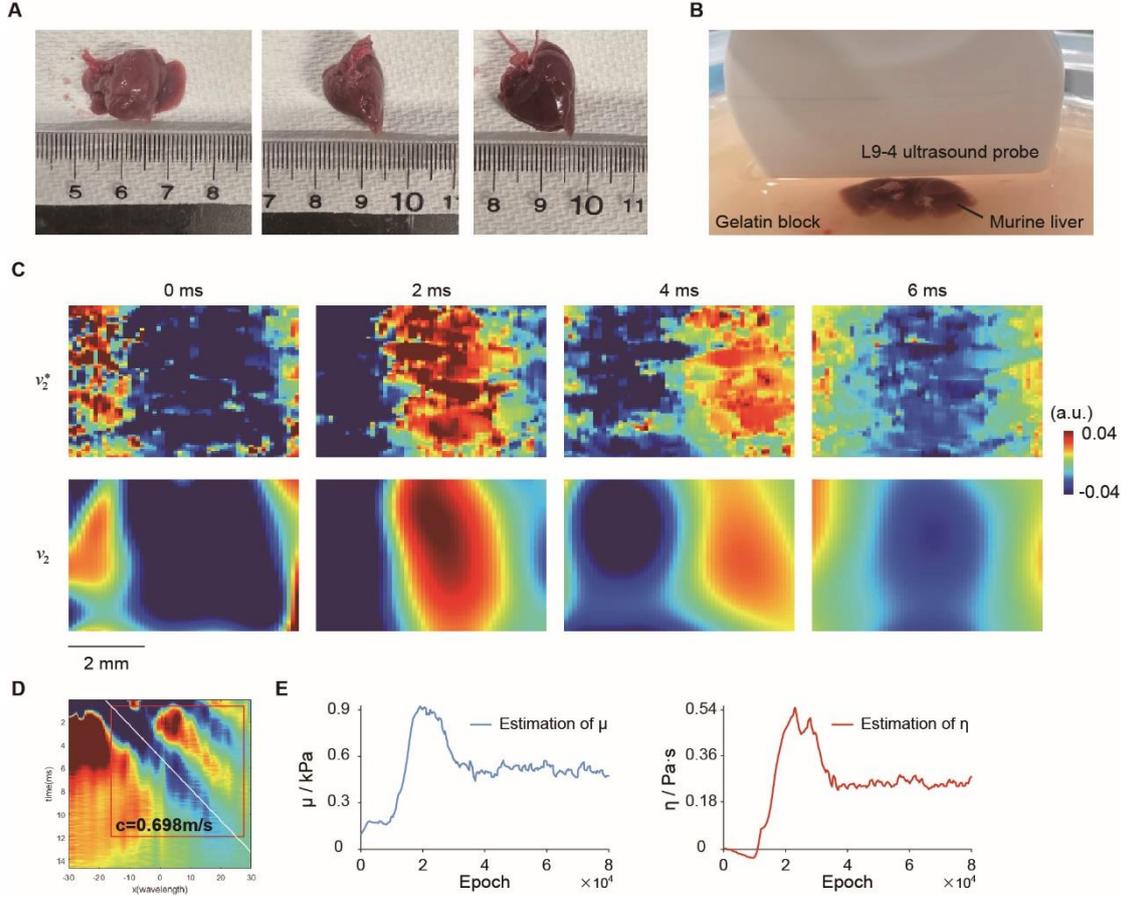

**Fig. 4. Inferring viscoelastic parameters of a small murine liver with SWVE-Net. (A)** Photographs of a murine liver sample with the thickness at sub-centimeter scale. **(B)** Experimental setup with the sample placed on a gelatin block under an ultrasound probe. **(C)** Vertical particle velocity field $v_2^*$ (first row) from experiments and predicted $v_2$ (second row) with SWVE-Net. **(D)** Representative spatio-temporal velocity field at a specific imaging depth, illustrating complex wave patterns. **(E)** Convergence of predicted viscoelastic parameters with SWVE-Net for a typical measurement.

Fig. 4C shows the vertical component of the particle velocity field ($v_2^*$), which serves as the experimental data for the SWVE-Net, along with the neural network's prediction of the corresponding shear wave field ($v_2$) subsequent to training. Notwithstanding the intricate interference patterns, the network precisely captured the overall shear wave dynamics. A representative spatio-temporal velocity field obtained at a specific imaging depth is depicted in Fig. 4D. Owing to the close proximity of the excitation position to the tissue boundary, reflected shear waves can be witnessed superimposed upon the primary wavefronts, which pose great challenges for conventional SWE methods relying on the traction of dispersion analysis to infer tissue viscoelastic parameters, but can be employed in SWVE-Net as multi-source data.

To evaluate the repeatability and robustness of the proposed methodology, three independent measurements were acquired at a consistent anatomical site and processed individually through SWVE-Net. The inferred viscoelastic parameters exhibited remarkable consistency across repeated measurements, with µ = 0.46 ± 0.02 kPa and η = 0.27 ± 0.01 Pa·s for a typical sample. The relatively small standard deviations demonstrate SWVE-Net's robustness in stably and precisely determining viscoelastic properties of small tissue samples, even under the complex wave propagation dynamics inherent to a real biological system.

**SWVE-Net enables quantitative assessment of viscoelastic properties of multi-organ types including brain, kidney and spleen in *ex vivo* experiments**

In addition to murine liver tissue, we extended our ex vivo experiments to soft tissues from multiple organs including porcine kidney, spleen, and brain tissues, further validating the cross-organ applicability of SWVE-Net. Similar to the shear wave imaging experiments of murine liver, focused acoustic radiation force was used to induce shear waves and the wave motion information was tracked with ultrafast ultrasound imaging. Representative experimental setups, including ultrasound probe positioning and corresponding tissue samples for each organ, are illustrated in Fig. 5A–C. Notably, these different organs may exhibit significantly distinct viscoelastic properties because of their different microstructures. The spatio-temporal velocity field derived from ultrafast ultrasound imaging, i.e., $v_2^*$, was used as the inputs of SWVE-Net to infer the viscoelastic parameters.

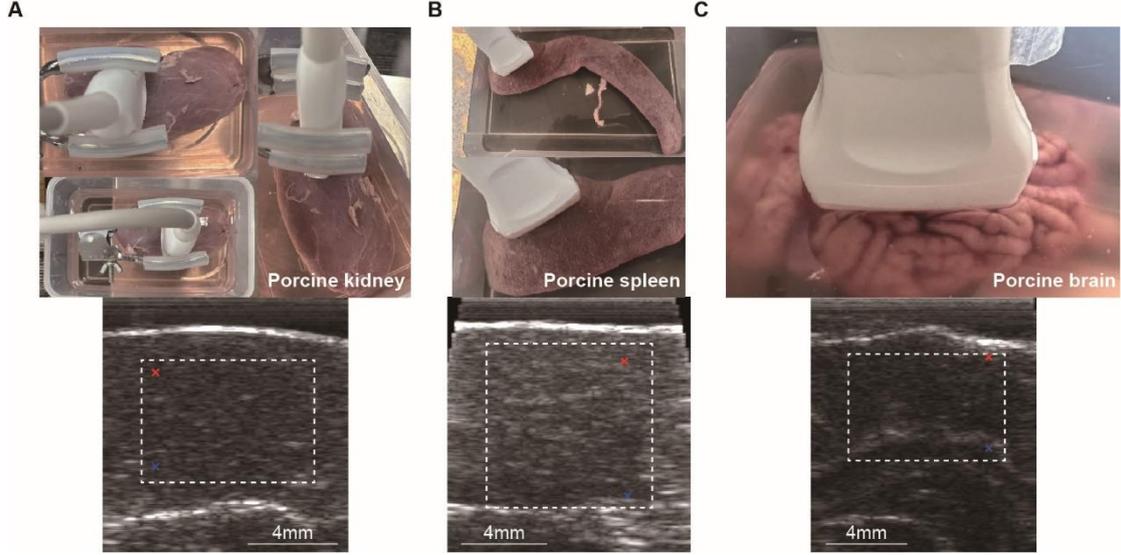

**Fig. 5.** Ex-vivo experiment setup (first row) and corresponding B-mode ultrasound images (second row) for multi-organ measurements. **(A)** Murine kidney **(B)** Murine spleen **(C)** Murine brain. White dashed squares denote the region of interest (ROI).

For every tissue type, three repeated measurements were performed at the same location and independently employed for viscoelastic property reconstruction through SWVE-Net. The quantitative inversion results are presented in Table 1. Besides the estimated shear modulus $\mu$ and viscosity $\eta$, we present the characteristic time ($\tau = \eta/\mu$), which is a key parameter defining the material's time-dependent response. The characteristic times for all tissues were of the order of $10^{-4}$ s (i.e., 0.1 ms), which is in accordance with the values reported in the literature [25-27].

**Table 1** Inversion results of viscoelastic parameters of different *ex vivo* samples

| *ex vivo* sample | $\mu$ / kPa | $\eta$ / (0.1Pa·s) | characteristic time / (0.1ms) |
| --- | --- | --- | --- |
| Murine liver | 0.46±0.02 | 2.74±0.12 | 5.96 |
| Porcine spleen | 5.26±0.53 | 20.02±1.04 | 3.80 |
| Porcine kidney | 5.96±0.05 | 15.95±0.16 | 2.68 |
| Porcine brain | 2.13±0.13 | 12.60±0.47 | 5.91 |

**SWVE-Net enables the quantification of viscoelastic properties of human living soft tissues**

Finally, to illustrate the potential clinical applicability of SWVE-Net in clinical scenarios, in vivo shear wave measurements were carried out on the biceps of a healthy male volunteer and the breast tissue of a healthy female volunteer. Figure 6A depicts B-mode ultrasound images of both the measured soft tissues, where the red and blue cross markers denote the start and end positions of the acoustic radiation force (ARF) excitation sequence. Taking the breast tissue as an example, Figure 6B and D show the vertical particle velocity $v_2^*$ measured in the in vivo experiment and the corresponding prediction made by SWVE-Net, respectively. As the shear wave propagates, the observed broadening of the shear wavefronts suggests viscoelastic dissipation within the tissue. The convergence feature of viscoelastic parameter estimation for the *in vivo* breast is presented in Figure 6C. As the loss function progressively diminishes during the training process, the predicted values of the shear modulus and viscosity attain stability, indicating the inferred parameters converge.

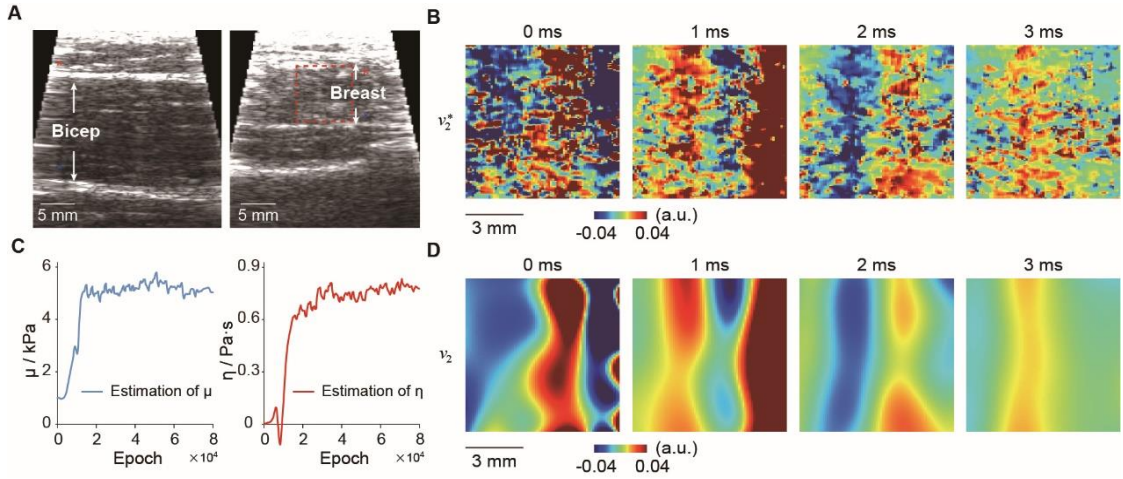

**Fig. 6. In vivo application of SWVE-Net on human living soft tissues. (A)** B-mode ultrasound images of the biceps and breast, red dashed square denotes the location and size of ROI. **(B)** Four typical snapshots of vertical particle velocities measured in breast tissue. **(C)** Convergence curves of predicted shear modulus and viscosity throughout the training epochs. **(D)** SWVE-Net predictions of vertical particle velocities.

Similar to ex vivo experiments, three repeated measurements were performed at the same location of the sample object. The reconstructed viscoelastic properties for *in vivo* tissues are summarized in Table 2. Notably, the entire reconstruction process was achieved without the need for extracting dispersion relations, which are difficult to be required in our in vivo experiments because of the relatively low SNRs of the wave motion signals. These scatters of identified results as shown in Table 2 demonstrate the effectiveness and translational potential of SWVE-Net for noninvasive, quantitative viscoelastic imaging in clinical settings.

**Table 2** Inversion results of viscoelastic parameters of different *in vivo* tissues

| *in vivo* sample | $\mu$ / kPa | $\eta$ / (0.1Pa·s) | Characteristic time / (0.1ms) |
|---|---|---|---|
| Breast (female) | 4.94±0.16 | 7.82±0.16 | 1.58 |
| Bicep (male) | 2.99±0.17 | 8.16±1.20 | 2.73 |

## Discussion

Tissue viscoelasticity offers significant insights into its diverse physiological and pathological states. Existing shear wave elastography (SWE) methods for inferring viscoelastic properties, especially the state-of-the-art approaches employed in clinical settings, depend on the determination of the dispersion relation of shear waves. These methods face difficulties in assessing viscoelastic parameters of small-sized or heterogeneous soft tissues due to the inherent challenges in discriminating between dual dispersion sources stemming from viscoelasticity and finite tissue dimensions. The SWVE-Net proposed in this study is a dispersion relation-free method. It takes particle velocities obtained from ultrasound imaging as inputs and offers a more direct and reliable approach to infer tissue viscoelastic properties, particularly in complex scenarios involving wave reflections and small tissue sizes. SWVE-Net incorporates viscoelastic wave equations into the loss function. This enables simultaneous forward modeling and constitutive parameter inference without the necessity of large training datasets. This characteristic renders it a more practical and efficient tool, considering that collecting extensive in vivo data can be arduous in a clinical environment.

Numerical simulations quantitatively verify the performance of SWVE-Net across various viscoelastic regimes. The results indicate that it can precisely measure the viscoelastic properties of a tissue sample as small as 5 millimeters in which wave reflection is significant. Extensive ex vivo experimental studies have been carried out to illustrate the utility of the proposed method in evaluating the viscoelasticity of a wide array of soft tissues, including murine liver, spleen, kidney, and brain tissues. The inferred characteristic viscoelastic constants of small liver tissues are reliable in multiple repeated measurements, and those obtained from ex vivo experiments are consistent with the values reported in the literature [25-27]. To showcase the potential clinical application of SWVE-Net, in vivo experiments were conducted on human living tissues, such as male biceps and female breast tissues. Our results reveal that SWVE-Net demonstrates consistent stability and robust performance in reconstructing both the elastic modulus and viscosity parameters of these human tissues.

Given that tissue viscoelasticity can act as a diagnostic biomarker for various diseases, SWVE-Net may have broad clinical applications. For instance, it could be utilized for grading the severity of hepatic lipid accumulation, detecting myocardial infarction boundaries, and differentiating malignant from benign tumors. In all these cases, determining the dispersion relation of shear waves is far from simple, and existing SWE methods may not be applicable. The dispersion-relation-free characteristics of SWVE-Net suggest that it may provide more accurate and reliable information about tissue viscoelasticity in these clinical situations. Moreover, SWVE-Net's ability to utilize multi-source data in the inversion of viscoelastic parameters is particularly valuable in clinics when reflection wave signals from different directions can be effectively employed, which may improve the accuracy of diagnosis.

To present the framework of SWVE-Net, we primarily focus on a specific viscoelastic model (Kelvin-Voigt model) in the present study. Living tissues may exhibit more complex viscoelastic behaviors that cannot be fully described by the KV model. In theory, integrating more sophisticated constitutive models into SWVE-Net is straightforward; however, this issue is outside the scope of this study and requires further exploration. The current ex vivo and in vivo experiments are restricted to several typical soft tissues from different species. Further studies are needed to test SWVE-Net on a broader range of tissues and in different patient populations. Additionally, although the method shows promise for clinical use, more systematic studies are required to establish guidelines in a clinical environment by using existing gold-standard diagnostic benchmarks as reference.

## Conclusion

In summary, we have proposed a framework and practical implementation for using physics-informed neural networks (PINNs) to infer viscoelastic properties (SWVE-Net) of soft tissues across multiple organs. The governing equations for wave motions in viscoelastic soft tissues were integrated into the loss functions of PINN, making this shear wave imaging methodology free from wave dispersion analysis and usable in diverse in vivo environments. We envision numerous clinical applications of SWE-Net such as for grading the severity of hepatic lipid accumulation, detecting myocardial infarction boundaries, and distinguishing between malignant and benign tumors. Aside from clinical use, SWVE-Net may find broad applications in tissue engineering. For instance, the viscoelastic properties of various soft materials employed in tissue engineering may evolve dynamically under in vivo conditions. Consequently, SWE-Net could be used to monitor the dynamic variation of time-dependent properties of these materials, assess the mechanical response of surrounding soft tissues to the foreign objects, and ultimately aid in the design of artificial soft materials that can restore tissue function.

## Materials and Methods

### Incremental Dynamics of Viscoelastic Materials

Here we consider the wave motion equation of viscoelastic materials, which is also the governing equation that needs to be embedded within the SWVE-Net framework. Our theoretical formulation follows the recently proposed incremental dynamics theory for pre-stressed viscoelastic solids [28], which models small-amplitude wave propagation superimposed on a static, finite pre-deformation. This framework is particularly relevant to soft tissue characterization, as biological tissues often exhibit both viscoelasticity and pre-existing stress under physiological or experimental conditions.

Briefly, we review the pre-stressed viscoelastic model [28] proposed in previous study. The general form of the incremental wave equation for the pre-stressed viscoelastic material is

$$-\widehat{p_0}_{,i} - G\hat{Q}_{,i} + G\mathcal{A}^0_{jikl}u_{l,jk} - \Omega\sigma^e_{Djk}u_{i,jk} = \rho u_{i,tt} \qquad (3)$$

where $i, j, k, l \in \{1, 2, 3\}$ represent the spatial coordinates—$(x_1, x_2, x_3)$. The subscript after the comma denotes the partial derivative with respect to the corresponding variable. $t$ denotes the time. $\widehat{p_0}$ denotes the increment of the Lagrange multiplier $p_0$ enforcing incompressibility, and $\hat{Q}$ is the increment of the volumetric part of the elastic stress $Q$ ($= \sigma^e_{ii}/3$, using Einstein summation over $i$). $\boldsymbol{\sigma}^e_D$ ($= \boldsymbol{\sigma}^e - Q$) denotes the deviatoric Cauchy stress. The tensor $\mathcal{A}^0_{jikl}$ is the fourth-order Eulerian elasticity tensor linearized around the prestressed state [29]. The two frequency-dependent parameters $G$ and $\Omega$ in Eq. (3) characterize the viscoelastic response:

$$G = 1 + \hat{\eta}(i\omega)^\alpha, \qquad \Omega = \hat{\eta}(i\omega)^\alpha \qquad (4)$$

where $\hat{\eta}$ (unit $s^\alpha$) and $\alpha \in (0,1)$ are viscoelastic parameters defined by the Kelvin-Voigt fractional derivative (KVFD) model [30]. $i$ in Eq. (4) denotes the imaginary unit. $\hat{\eta}$ denotes the ratio of material viscosity to elasticity, and $\alpha$ is a fractional order. When $\alpha = 0$, the model reduces to elastic behavior; when $\alpha = 1$, it recovers to the viscoelastic Kelvin-Voigt model.

In the present study, we take the linear viscoelastic Kelvin-Voigt model without pre-stress for illustration, i.e.,

$$G = 1 + \hat{\eta}i\omega, \quad \Omega = \hat{\eta}i\omega, \quad \boldsymbol{\sigma}^e_D = Q = 0, \quad \mathcal{A}^0_{jikl} = C_{jikl} \qquad (5)$$

where $C_{jikl}$ denotes the fourth-order elasticity tensor for the incompressible material. Inserting Eq. (5) into Eq. (3) yields the wave equation for linear viscoelastic materials:

$$-p_{0,i} + (1 + \hat{\eta} i\omega) C_{jikl} u_{l,jk} = \rho u_{i,tt} \tag{6}$$

For isotropic materials, it holds that $C_{jikl} u_{l,jk} = \mu u_{i,jj}$. Furthermore, by replacing the harmonic term $i\omega$ with its equivalent temporal derivative, Eq. (6) simplifies to:

$$\rho u_{i,tt} = -p_{0,i} + \mu u_{i,jj} + \eta \dot{u}_{i,jj} \tag{7}$$

where $\dot{u}_i\ (= \partial u_i/\partial t)$ denotes the particle velocity $v_i$, and $\eta\ (= \hat{\eta}\mu)$ denotes the viscosity of the dashpot. Here, we take the time derivative of Eq. (7) to get the wave equation expressed in terms of particle velocity, which is frequently used in SWE. For wave motions in $x_1 - x_2$ plane, the wave equations can be written as

$$\rho v_{i,tt} = -p_{,i} + \mu v_{i,jj} + \eta \dot{v}_{i,jj}, \quad i \in \{1,2\} \tag{8}$$

where $p = \frac{\partial p_0}{\partial t}$, $p_{,i} = \frac{\partial p}{\partial x_i}$, $\dot{v}_i = \frac{\partial v_i}{\partial t}$. And the incompressibility constraint can be written as

$$v_{1,1} + v_{2,2} = 0. \tag{9}$$

Here we introduce a stream function $\psi = \psi(x_1, x_2, t)$ where $(v_1, v_2) = (\frac{\partial \psi}{\partial x_2}, -\frac{\partial \psi}{\partial x_1})$, which automatically satisfied the incompressible constraints. Therefore, the wave motions are fully described by the spatiotemporal field $\psi$ and $p$, while the viscoelastic property $\mu$ and $\eta$ are defined as global uniform parameters.

**Training Configuration**

The SWVE-Net architecture and hyperparameters were configured as follow. The fully connected feed-forward architecture was used, as suggested in previous studies on PINN [22, 24, 31], and hyperbolic tangent function (tanh) was utilized as the activation function [32]. The input variables and output variables are set to 3 and 2, respectively; while the hidden layers and neurons per layer are set to 8 and 20, respectively. For the loss function, the weights of the data term relative to the PDE term ($\lambda_{Data}/\lambda_{PDE}$) were set to $10^{-1}$ for simulation datasets and $10^{-4}$ for experimental datasets. Training data points were sampled in the region of interest (ROI), with batch sizes of $M = 1 \times 10^4$ across all datasets. The SWVE-Net model was implemented using TensorFlow 2.10.0 (CUDA 11.2, cuDNN 8.1) and trained on a system equipped with two NVIDIA GeForce RTX 4090 GPUs. On a single GPU, each 100 training epochs required approximately 7.5 seconds, with the total training time to achieve reliable inversion results being around 100 minutes.

**Finite element simulations**

To validate the proposed method, finite element simulations were carried out using Abaqus/Standard (Version 6.14, Dassault Systems, Waltham, MA, USA). We constructed two simulation models: a two-dimensional square domain large enough to preclude wave reflections at the boundaries, and a small-dimension model depicting a cross-sectional view of a murine liver to mirror the experimental environment, in which substantial wave reflections and dispersion occur. The boundaries of the small-dimension domain were meticulously delineated based on the 2D ultrasound images of a murine liver tested in the subsequent ex vivo experiments. The viscoelastic behavior of the material was modeled by employing the classical Voigt model. A user-defined material subroutine (UMAT) was developed to implement the constitutive behavior. Material properties were uniformly assigned across the domain, with a mass density of 1,000 kg/m³, a shear modulus of 4 kPa, and a Poisson's ratio of 0.4999. To illustrate the method's efficacy across diverse viscoelastic regimes, two distinct viscosity coefficients were taken into account: $\eta_1$= 0.15 Pa·s and $\eta_1$= 1.5 Pa·s in the first finite element model. For the small-dimension model to simulate the murine liver, the viscoelastic properties were set as µ = 0.5 kPa and η = 0.25 Pa·s.

The simulations were executed in the time domain via a dynamic implicit procedure. Two analysis steps were defined to model the processes of shear wave generation and propagation, with durations of 0.1 ms and 5 ms, respectively. In the first step, a Gaussian-shaped body force e f was applied.

$$\boldsymbol{f} = \boldsymbol{f}_0 \exp\left[-\frac{(x_1 - x_1^{(i)})^2}{2r_1^2} - \frac{(x_2 - x_2^{(i)})^2}{2r_2^2}\right] \quad (10)$$

where $\boldsymbol{f}_0 = [0, f_0]$ denotes the direction and peak magnitude of the force ($f_0$= 0.001 N/ mm²) of the peak body force. This force field emulates the acoustic radiation force (ARF) generated by an ultrasound beam, with the focal point coordinates $(x_1^{(i)}, x_2^{(i)})$ ($i$ = 1, 2, …, 10) moving along the vertical axis at a speed of 40 Mach. The spatial distribution parameters were taken as $r_1$ = 0.5 mm and $r_2$ = 1.1 mm, consistent with experimental conditions performed subsequently. In the second step, the propagation of the generated shear waves was simulated, and the particle velocity $v_2^*$ was recorded at intervals of 0.1 ms. The ARFs were applied on the left side of the field of view (FOV), generating shear waves that propagated through the region of interest (ROI) towards the right-hand side.

The computational domain was meshed using CPE8RH elements (8-node biquadratic plane strain elements with hybrid formulation and reduced integration), with an approximate element size of 0.05 × 0.05 mm². Convergence was confirmed by comparing results with those of a refined mesh. For subsequent inverse analysis, the results were linearly interpolated onto a uniform grid of 0.05 × 0.1 mm² and input into the SWVE-Net for viscoelastic property reconstruction.

### *Ex vivo* sample preparation and experimental setup

*Ex-vivo* tissue specimens across multiple-organs, encompassing porcine kidneys, porcine spleens, porcine brains, and murine livers, were collected on the day of animal euthanasia. Throughout the experiments, all specimens were placed within a container filled with physiological saline and maintained at room temperature to preserve tissue hydration. The ultrasound probe was positioned without direct contact with the specimens to prevent the application of additional mechanical pressure. Specifically, for murine livers, blood vessels were ligated immediately upon extraction to ensure optimal ultrasound imaging quality. Subsequently, the specimens were placed on a gelatin block to minimize strong reflections from the container's bottom. Shear wave elastography (SWE) measurements were carried out using a Verasonics Vantage 64LE system (Verasonics Inc., Kirkland, WA, USA) equipped with an L9-4 linear array transducer (central frequency: 6.5 MHz; 128 elements; element pitch: 0.3 mm). Shear waves were generated by focused acoustic radiation forces (ARFs) utilizing 32 elements (P5 voltage approximately 25 V; aperture size approximately 10 mm; uniform apodization). Six focused beams were sequentially applied within the ex-vivo specimens, generating a moving ARF that propagated along the depth direction. ARFs were applied to either the left or right side of the field of view (FOV). After ARF excitation, the system transitioned to plane wave imaging mode to capture shear wave propagation. In this mode, all 128 elements were employed for transmission (P5 voltage approximately 25 V; aperture size approximately 40 mm; uniform apodization), while only the 64 central elements were used for reception. The imaging frame rate was set at 10 kHz. In - phase and quadrature (IQ) data were acquired and processed offline to estimate the particle velocity using Loupas' phase estimator method [33].

*In vivo* **measurements**

To elucidate the potential clinical applicability of SWVE-Net, shear wave imaging was conducted on the biceps brachii of a healthy male volunteer and the breast tissue of healthy female volunteer. The experimental protocol received approval from the institutional review board at Tsinghua University (ethical approval number: Tsinghua UNIV. 20210039). During the measurement process, volunteers were instructed to place the measured tissue in a horizontal orientation on the platform. They were required to keep their muscles in a relaxed state and maintain a steady breathing pattern. These measures were taken to minimize motion-induced artifacts during the acquisition of shear waves. When performing shear wave elastography (SWE), the target tissue was precisely positioned at the center of the region of interest (ROI). The sizes of the ROIs for viscoelastic inversion in the biceps brachii and breast tissue were 10×5 mm² and 8×8 mm², respectively. Acoustic radiation forces (ARFs) were applied to either the left or right side of the ROI, and each measurement had a duration of 4 ms. The experimental system and procedure employed in this *in-vivo* study were consistent with those utilized in the *ex-vivo* experiments. Particle velocity was recorded, and SWVE-Net was used to infer the shear modulus of the biceps brachii muscle and the breast tissue.

# References

[1] F. Català-Castro, S. Ortiz-Vásquez, C. Martínez-Fernández, F. Pezzano, C. Garcia-Cabau, M. Fernández-Campo, N. Sanfeliu-Cerdán, S. Jiménez-Delgado, X. Salvatella, V. Ruprecht, Measuring age-dependent viscoelasticity of organelles, cells and organisms with time-shared optical tweezer microrheology, Nature Nanotechnology (2025) 1-10.

[2] S. Hurst, B.E. Vos, M. Brandt, T. Betz, Intracellular softening and increased viscoelastic fluidity during division, Nature Physics 17(11) (2021) 1270-1276.

[3] O. Chaudhuri, J. Cooper-White, P.A. Janmey, D.J. Mooney, V.B. Shenoy, Effects of extracellular matrix viscoelasticity on cellular behaviour, Nature 584(7822) (2020) 535-546.

[4] Y. Wu, Y. Song, J. Soto, T. Hoffman, X. Lin, A. Zhang, S. Chen, R.N. Massad, X. Han, D. Qi, Viscoelastic extracellular matrix enhances epigenetic remodeling and cellular plasticity, Nature Communications 16(1) (2025) 4054.

[5] A. Patel, H.O. Lee, L. Jawerth, S. Maharana, M. Jahnel, M.Y. Hein, S. Stoynov, J. Mahamid, S. Saha, T.M. Franzmann, A liquid-to-solid phase transition of the ALS protein FUS accelerated by disease mutation, Cell 162(5) (2015) 1066-1077.

[6] W. Fan, K. Adebowale, L. Váncza, Y. Li, M.F. Rabbi, K. Kunimoto, D. Chen, G. Mozes, D.K.-C. Chiu, Y. Li, Matrix viscoelasticity promotes liver cancer progression in the pre-cirrhotic liver, Nature 626(7999) (2024) 635-642.

[7] J. Chen, J.A. Talwalkar, M. Yin, K.J. Glaser, S.O. Sanderson, R.L. Ehman, Early detection of nonalcoholic steatohepatitis in patients with nonalcoholic fatty liver disease by using MR elastography, Radiology 259(3) (2011) 749-756.

[8] C. Pislaru, M.W. Urban, S.V. Pislaru, R.R. Kinnick, J.F. Greenleaf, Viscoelastic properties of normal and infarcted myocardium measured by a multifrequency shear wave method: comparison with pressure-segment length method, Ultrasound in medicine & biology 40(8) (2014) 1785-1795.

[9] I. Sack, Magnetic resonance elastography from fundamental soft-tissue mechanics to diagnostic imaging, Nature Reviews Physics 5(1) (2023) 25-42.

[10] R. Muthupillai, D. Lomas, P. Rossman, J.F. Greenleaf, A. Manduca, R.L. Ehman, Magnetic resonance elastography by direct visualization of propagating acoustic strain waves, science 269(5232) (1995) 1854-1857.

[11] A.P. Sarvazyan, O.V. Rudenko, S.D. Swanson, J.B. Fowlkes, S.Y. Emelianov, Shear wave elasticity imaging: a new ultrasonic technology of medical diagnostics, Ultrasound in medicine & biology 24(9) (1998) 1419-1435.

[12] K. Nightingale, S. McAleavey, G. Trahey, Shear-wave generation using acoustic radiation force: in vivo and ex vivo results, Ultrasound in medicine & biology 29(12) (2003) 1715-1723.

[13] J.-L. Gennisson, M. Rénier, S. Catheline, C. Barrière, J. Bercoff, M. Tanter, M. Fink, Acoustoelasticity in soft solids: Assessment of the nonlinear shear modulus with the acoustic radiation force, The Journal of the Acoustical Society of America 122(6) (2007) 3211-3219.

[14] P. Song, M.C. Macdonald, R.H. Behler, J.D. Lanning, M.H. Wang, M.W. Urban, A. Manduca, H. Zhao, M.R. Callstrom, A. Alizad, Two-dimensional shear-wave elastography on conventional ultrasound scanners with time-aligned sequential tracking (TAST) and comb-push ultrasound shear elastography (CUSE), IEEE transactions on ultrasonics, ferroelectrics, and frequency control 62(2) (2015) 290-302.

[15] J. Ormachea, K.J. Parker, Comprehensive viscoelastic characterization of tissues and the inter-relationship of shear wave (group and phase) velocity, attenuation and dispersion, Ultrasound in Medicine & Biology 46(12) (2020) 3448-3459.

[16] N.C. Rouze, Y. Deng, C.A. Trutna, M.L. Palmeri, K.R. Nightingale, Characterization of viscoelastic materials using group shear wave speeds, IEEE transactions on ultrasonics, ferroelectrics, and frequency control 65(5) (2018) 780-794.

[17] X. Chen, X. Li, S. Turco, R.J. Van Sloun, M. Mischi, Ultrasound viscoelastography by acoustic radiation force: A state-of-the-art review, IEEE Transactions on Ultrasonics, Ferroelectrics, and Frequency Control (2024).

[18] I.Z. Nenadic, M.W. Urban, H. Zhao, W. Sanchez, P.E. Morgan, J.F. Greenleaf, S. Chen, Application of attenuation measuring ultrasound shearwave elastography in 8 post-transplant liver patients, 2014 IEEE International Ultrasonics Symposium, IEEE, 2014, pp. 987-990.

[19] N.C. Rouze, M.L. Palmeri, K.R. Nightingale, An analytic, Fourier domain description of shear wave propagation in a viscoelastic medium using asymmetric Gaussian sources, The Journal of the Acoustical Society of America 138(2) (2015) 1012-1022.

[20] N.C. Rouze, Y. Deng, M.L. Palmeri, K.R. Nightingale, Accounting for the spatial observation window in the 2-D Fourier transform analysis of shear wave attenuation, Ultrasound in medicine & biology 43(10) (2017) 2500-2506.

[21] P. Kijanka, L. Ambrozinski, M.W. Urban, Two point method for robust shear wave phase velocity dispersion estimation of viscoelastic materials, Ultrasound in medicine & biology 45(9) (2019) 2540-2553.

[22] M. Raissi, P. Perdikaris, G.E. Karniadakis, Physics-informed neural networks: A deep learning framework for solving forward and inverse problems involving nonlinear partial differential equations, Journal of Computational physics 378 (2019) 686-707.

[23] Abaqus User Subroutines Reference Guide (6.14). http://62.108.178.35:2080/v6.14/books/sub/default.htm?startat=ch01s01asb09.html.

[24] Z. Yin, G.-Y. Li, Z. Zhang, Y. Zheng, Y. Cao, SWENet: a physics-informed deep neural network (PINN) for shear wave elastography, IEEE Transactions on Medical Imaging 43(4) (2023) 1434-1448.

[25] X. Chen, Y. Shen, Y. Zheng, H. Lin, Y. Guo, Y. Zhu, X. Zhang, T. Wang, S. Chen, Quantification of liver viscoelasticity with acoustic radiation force: a study of hepatic fibrosis in a rat model, Ultrasound in medicine & biology 39(11) (2013) 2091-2102.

[26] S. Chen, M.W. Urban, C. Pislaru, R. Kinnick, J.F. Greenleaf, Liver elasticity and viscosity

quantification using shearwave dispersion ultrasound vibrometry (SDUV), 2009 Annual International Conference of the IEEE Engineering in Medicine and Biology Society, IEEE, 2009, pp. 2252-2255.

[27] C. Amador, M.W. Urban, S. Chen, J.F. Greenleaf, Shearwave dispersion ultrasound vibrometry (SDUV) on swine kidney, IEEE transactions on ultrasonics, ferroelectrics, and frequency control 58(12) (2011) 2608-2619.

[28] Y. Jiang, G.-Y. Li, Z. Zhang, S. Ma, Y. Cao, S.-H. Yun, Incremental dynamics of prestressed viscoelastic solids and its applications in shear wave elastography, International Journal of Engineering Science 215 (2025) 104310.

[29] R.W. Ogden, Incremental statics and dynamics of pre-stressed elastic materials, Waves in nonlinear pre-stressed materials, Springer2007, pp. 1-26.

[30] A. Capilnasiu, L. Bilston, R. Sinkus, D. Nordsletten, Nonlinear viscoelastic constitutive model for bovine liver tissue, Biomechanics and modeling in mechanobiology 19 (2020) 1641-1662.

[31] M. Rasht-Behesht, C. Huber, K. Shukla, G.E. Karniadakis, Physics-informed neural networks (PINNs) for wave propagation and full waveform inversions, Journal of Geophysical Research: Solid Earth 127(5) (2022) e2021JB023120.

[32] K. Hornik, M. Stinchcombe, H. White, Multilayer feedforward networks are universal approximators, Neural networks 2(5) (1989) 359-366.

[33] T. Loupas, J. Powers, R.W. Gill, An axial velocity estimator for ultrasound blood flow imaging, based on a full evaluation of the Doppler equation by means of a two-dimensional autocorrelation approach, IEEE transactions on ultrasonics, ferroelectrics, and frequency control 42(4) (2002) 672-688.